# First Results for Solar Soft X-ray Irradiance Measurements from the Third Generation Miniature X-Ray Solar Spectrometer

Thomas N. Woods[1,2], Bennet Schwab[2], Robert Sewell[2], Anant Kumar Telikicherla Kandala[3], James Paul Mason[4], Amir Caspi[5], Thomas Eden[2], Amal Chandran[2], Phillip C. Chamberlin[2], Andrew R. Jones[2], Richard Kohnert[2], Christopher S. Moore[6], Stanley C. Solomon[7], Harry Warren[8]

## Abstract

Three generations of the Miniature X-ray Solar Spectrometer (MinXSS) have flown on small satellites with the goal "to explore the energy distribution of soft X-ray (SXR) emissions from the quiescent Sun, active regions, and during solar flares, and to model the impact on Earth's ionosphere and thermosphere". The primary science instrument is the Amptek X123 X-ray spectrometer that has improved with each generation of the MinXSS experiment. This third generation MinXSS-3 has higher energy resolution and larger effective area than its predecessors and is also known as the Dual-zone Aperture X-ray Solar Spectrometer (DAXSS). It was launched on the INSPIRESat-1 satellite on 2022 February 14, and INSPIRESat-1 has successfully completed its 6-month prime mission. The INSPIRESat-1 is in a dawn-dusk, Sun-Synchronous Orbit (SSO) and therefore has 24-hour coverage of the Sun during most of its mission so far. The rise of Solar Cycle 25 (SC-25) has been observed by DAXSS. This paper introduces the INSPIRESat-1 DAXSS solar SXR observations, and we focus the science results here on a solar occultation experiment and multiple flares on 2022 April 24. One key flare result is that the reduction of elemental abundances appears greatest during the flare impulsive phase and thus highlights the important role of chromospheric evaporation during flares to inject warmer plasma into the coronal loops. Furthermore, these results are suggestive that the amount of chromospheric evaporation is related to flare temperature and intensity.

---

[1] Corresponding author  tom.woods@lasp.colorado.edu     *Version 2023-07-29*
[2] University of Colorado (CU) Laboratory for Atmospheric and Space Research (LASP), Boulder, CO
[3] Department of Electrical and Computer Engineering, University of Alberta, Edmonton, Canada
[4] The Johns Hopkins University (JHU) Applied Physics Laboratory (APL), Laurel, MD
[5] Southwest Research Institute (SwRI), Boulder, CO
[6] Center for Astrophysics, Harvard & Smithsonian, Boston, MA
[7] National Center for Atmospheric Research (NCAR) High Altitude Observatory (HAO), Boulder, CO
[8] Naval Research Laboratory (NRL), Washington, DC

1. Introduction

The solar soft X-ray (SXR: 0.1-10 nm) and extreme ultraviolet (EUV: 10-120 nm) emissions are highly variable and have been studied for the past five decades to understand the solar corona, transition region, and upper chromosphere (e.g., Aschwanden 2019). The new SXR observations as introduced here have potential to further address the outstanding issues about coronal heating as related to quiet-Sun plasma elemental abundances, active region formation and evolution, and flare energetics (e.g., Del Zanna & Mason 2018). The variability of the solar SXR and EUV emissions is also important for terrestrial and planetary upper atmospheres studies because these solar emissions have direct influence on those atmospheres, such as heating the thermosphere, creating the ionosphere, and driving changes on many different time scales ranging from minutes by flares to years by the 11-year solar activity cycle (e.g., Ward et al. 2021). Broadband solar SXR measurements, such as from the Geostationary Operational Environmental Satellite (GOES) X-Ray Sensor (XRS) (Garcia 1994), have been made nearly continuously for the past four decades, and they have been valuable as a monitor for solar coronal activity and resolving some differences in terrestrial ionosphere models and measurements. However, the lack of spectral resolution in the SXR range has led to gaps in understanding the terrestrial photoelectron production and related impacts in the ionosphere D and E layers. Furthermore, the lack of spectral measurements in the SXR range has limited the physical understanding of the corona emissions in the solar SXR spectrum; those emissions include the many emission lines (bound-bound) and the thermal radiative recombination (free-bound) and thermal and non-thermal bremsstrahlung (free-free) continua. Using recent advances in satellite and instrument miniaturization, the NASA-funded Miniature X-ray Solar Spectrometer (MinXSS) mission (Mason et al. 2016) was developed to make new solar SXR spectral observations from CubeSats and to begin to address those issues. MinXSS demonstrated the quality of pioneering science available even from such a small and inexpensive platform (e.g., Spence et al. 2022).

The MinXSS CubeSat project includes the MinXSS-1 CubeSat deployed from the International Space Station (ISS) in May 2016 for a 1-year mission, the MinXSS-2 CubeSat mission with a launch in December 2018, and a flight of opportunity for MinXSS-3 payload on the small satellite INSPIRESat-1 with a launch in February 2022. In all three missions, the primary science instrument is the Amptek X123 Silicon Drift Detector (SDD) with a Be foil filter to provide solar soft X-ray (SXR) irradiance (integrated full-disk) measurements for the 0.5-20 keV (0.06-2.5 nm) range. Each generation of the MinXSS experiment has incorporated advances for the X123 to improve its energy resolution and dynamic range. The MinXSS-1 X123 had an energy resolution of 0.22 keV at 2 keV (Mason et al. 2016; Moore et al. 2016) and made daily observations until its reentry on 2017 May 6 (Woods et al. 2017; Moore et al. 2018). The MinXSS-2 flew an upgraded version of the X123 called the Fast SDD, which has wider dynamic range and lower noise than for MinXSS-1 and consequently has an improved energy resolution of 0.14 keV at 2 keV (Mason et al. 2020; Moore et al. 2018). The MinXSS-2 CubeSat was deployed from Spaceflight Industry's SSO-A multi-satellite launch on a Falcon-9 on 2018 December 3. MinXSS-2 made good solar observations during mostly quiet-sun periods, but the MinXSS-2 mission ended unexpectedly early on 2019 January 7 when a single event latchup (SEL) event appears to have permanently damaged the memory storage card on its Command and Data Handling (CDH) board and thus halted data operations for MinXSS-2.

With this unexpected anomaly, the MinXSS-3 instrument was quickly developed and delivered in 2019 as a flight-of-opportunity payload for the INSPIRESat-1 mission (Chandran et al. 2021); however, the launch of the INSPIRESat-1 on the Indian Space Research Organization's Polar Space Launch Vehicle C was delayed by two years due to the world-wide COVID19 pandemic. The MinXSS-3 payload includes flight-spare boards from MinXSS-2 and a technology-demonstration version of the X123 Fast SDD using a novel aperture design, named the Dual-zone Aperture X-ray Solar Spectrometer (DAXSS) (Schwab et al. 2020). The Amptek advances for its X123 Fast SDD and improvements in noise reduction on INSPIRESSat-1 compared to the MinXSS buses provide improved energy resolution of 0.09 keV at 2 keV for DAXSS. This DAXSS unit flew successfully on a 2018 June 18 sounding rocket flight (Schwab et al. 2020) and was integrated into the MinXSS-3 payload in 2019. The 60% improvement in energy resolution for each MinXSS generation enables better emission feature identification and effectively improves the accuracy of MinXSS science results as demonstrated later.

In relation to other solar irradiance observations, the MinXSS-1 solar SXR spectral measurements in the 0.5-10 keV range in 2016-2017 have helped to fill the gap between hard X-ray (HXR: 10–1000 keV) observations from the Reuven Ramaty High Energy Solar Spectroscopic Imager (RHESSI, Lin et al. 2002) and the EUV (0.01-0.2 keV) observations from the Solar Dynamics Observatory (SDO, Pesnell et al. 2012). In addition, there are other recent solar SXR spectral measurements that when combined with the MinXSS observations provide improved temporal and/or spectral coverage over Solar Cycles 24 and 25. Those include CORONAS-PHOTON SphinX (Gburek et al. 2011; Sylwester et al. 2012), MErcury Surface, Space ENvironment, Geochemistry, and Ranging (MESSENGER) Solar Assembly for X-rays (SAX) (Schlemm et al. 2007; Dennis et al. 2015), Chandrayaan-2 Solar X-ray Monitor (XSM) (Mithun et al. 2020), and Solar Orbiter Spectrometer/Telescope for Imaging X-rays (STIX) (Krucker et al. 2020). MinXSS complements these other solar SXR spectral measurements by observing at lower energy and with improved energy resolution. Despite having several solar SXR spectrometers observing during Solar Cycle 24 and 25, many of the missions do not overlap in time and so there continues to be many time gaps that hamper more detailed studies for solar and ionosphere-thermosphere research.

This paper focuses on the improved solar SXR spectral measurements by DAXSS aboard the INSPIRESat-1 small satellite. As described in more detail by Schwab et al. (2020), the DAXSS uses the improved 2018 version of Amptek's X123 Fast SDD, which provides the advances in the energy resolution, and DAXSS also incorporates a novel approach for the aperture with two zones to improve the sensor response for energies higher than 1.5 keV. The small-aperture zone behaves the same as previous MinXSS instruments with a Be foil filter, and the large-aperture zone has a much larger aperture area with addition of a Kapton thin-film filter above the Be filter. As illustrated in Figure 1, the DAXSS effective area is 600 times larger than the MinXSS-1 effective area for energies above 4 keV. This significant improvement in effective area enables DAXSS to measure the solar spectra much more accurately (due to having much larger signal), but this does have a consequence in that DAXSS solar observations are limited to flare levels below M3 (due to the detector saturation limit for higher count rates). The response enhancements at energies below 1 keV are because the DAXSS Be filter is thinner than that for MinXSS-1. Figure 1 also shows the improvement in energy resolution, being a factor of 2.9 better for DAXSS than MinXSS-1 at 1 keV and a factor of 1.7 better at 7 keV.

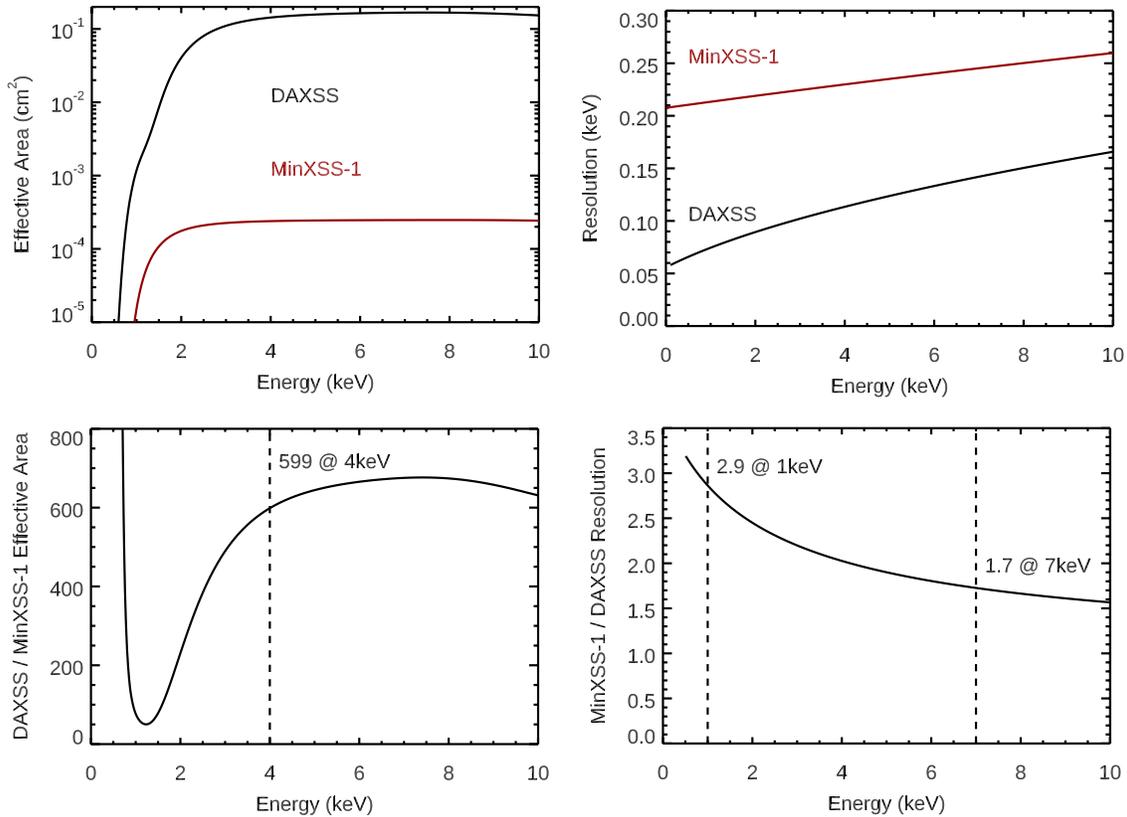

Figure 1. As compared to MinXSS-1, the DAXSS sensor has significant improvements in effective area (left panels) and energy resolution (right panels).

MinXSS-3 includes two other solar sensors: Solar Position Sensor (SPS) and the Pico Spectral Irradiance Monitor (PicoSIM). The SPS is a quadrant Si photodiode (Luna SD-085-23-21-021) to provide solar position information with arc-sec level precision; however, it has had saturated solar signals during normal solar observations. During the transition to orbit eclipse (solar occultation), the SPS signal is not saturated, and so the SPS can provide more useful solar position information during solar occultation measurements. The PicoSIM is the AMS AS7263 IC that has six CMOS sensors with spectral passbands in the visible and near-infrared range (610-860 nm with ~20 nm FWHM bandpass). Because the PicoSIM internal filters are interference filters that have spectral shifts with varying angle of incidence, the PicoSIM signals vary significantly with spacecraft pointing angle, but PicoSIM signals are stable when INSPIRESat-1 is in its normal science mode. We are analyzing these PicoSIM data as a means to provide solar position information, but with expected precision much less than what the SPS is capable of providing if the SPS signal was not saturated.

## 2. DAXSS Solar Observations and Data Products

The INSPIRESat-1 normal science observations are to view the Sun with DAXSS during orbit sunlit periods and to orient into the spacecraft ram direction for the Compact Ionospheric Probe (CIP) ionospheric plasma measurements during orbit eclipse periods (Chandran et al. 2021), though due to the orbit geometry, eclipses are rare. The MinXSS-3 supporting electronics stay powered on during orbit eclipse, but the 5-W X123 unit is turned off to conserve power during eclipse.

During early-orbit commissioning and in case of an anomaly, INSPIRESat-1 is in its Safe mode, or if the battery voltage is too low, its Phoenix mode. The Safe mode is solar oriented with a few degrees stability, and so DAXSS with its wide 8° full field of view also makes good solar observations even in Safe mode.

With INSPIRESat-1 in approximately circular polar orbit with altitude of about 540 km, there are about fifteen orbits per day. The orbit eclipse periods have, so far, been limited from 2022 May 1 (Day Of Year, DOY, 121) to 2022 August 11 (DOY 223). Therefore, it has been possible to have DAXSS on and making solar observations 24-hours per day for much of the mission. In this configuration, the solar zenith angle relative to the Earth-satellite line is near 90°, but there are times in some orbits when the solar zenith angle exceeds 90° with a view through Earth's atmosphere, and thus DAXSS sometimes provides solar occultation measurements, which can be analyzed to provide thermospheric density results as shown later.

The normal integration time for DAXSS is 9 seconds, and the MinXSS-3 science packet at this cadence includes the DAXSS X123 spectra, PicoSIM six-channel photometer data, SPS data, and MinXSS-3 housekeeping data (voltages, currents, temperatures). The INSPIRESat-1 beacon (housekeeping) packets do not provide pointing information, so we had planned to use the SPS results to provide accurate solar position information. As mentioned earlier, the SPS solar signal is saturated and so the solar position information is not routinely known. However, we get confirmation of good solar pointing periods when the PicoSIM signals are stable, as is seen during INSPIRESat-1 Normal Science mode. INSPIRESat-1 uses the Blue Canyon Technology (BCT) Attitude Determination and Control System (ADCS) as used for MinXSS-1 and -2 CubeSats, so we are currently assuming that INSPIRESat-1 is getting similar ~10-arc-sec (1-s) pointing for the solar observations as achieved for MinXSS-1 (Mason et al. 2017). As an on-going validation effort, we are working on analysis of the INSPIRESat-1 ADCS data packets and also on relating the ADCS information to the PicoSIM signal variations when INSPIRESat-1 is in its Safe mode. We plan to include the estimated solar pointing information derived from PicoSIM signal variation in future DAXSS data products.

The DAXSS response with incidence angle is very flat (Schwab et al. 2020), so DAXSS is making good solar observations during both INSPIRESat-1 Normal Science and Safe modes. Therefore, the DAXSS data processing includes solar SXR spectra from both modes in its Level 1 science product. The Level 1 processing algorithm assigns the photon energy for each of its 1024 bins and folds in the DAXSS response based on radiometric calibrations at NIST SURF (Schwab et al. 2020). The relationship of the X123 spectral bins to energy was determined pre-flight using laboratory radioactive sources and is validated in-flight by the position of key solar emission features in the SXR spectra (Schwab et al. 2020). There are two corrections made to the raw spectra before folding in the response function. One is a correction for X123 photon-counting deadtime based on total signal and preflight linearity calibration (Moore et al. 2018; Schwab et al. 2020), and the other is a small correction for particle background. During intense geomagnetic storms, the DAXSS background level goes up from essentially zero to a few counts per spectral bin per 9-sec integration period. The spectral slope of this particle background is pretty flat; nonetheless, a linear trend over energy is fit using the 12-20 keV signals and applied to all energies. After folding in the response function to get irradiance units for the solar SXR spectra, a correction (scaling factor) to 1-AU solar distance is applied. The Level 1 processing does exclude DAXSS spectra if the solar tangent ray height is below 300 km (occultation / orbit eclipse), if the X123 total signal is near its saturation level (above about M3 flare level), or if the X123 sensor temperature is not stable (for ~1 minute after X123 is turned on). The Level 1 product includes the DAXSS spectra with its time, energy, irradiance, precision, accuracy, and correction factors, as well as satellite orbit information.

Most of the DAXSS solar spectra can be grouped into (1) quiescent sun (non-flaring), (2) flares as discussed in Sections 4 and 5, and (3) occultation data (atmospheric absorption) as discussed in Section 3. Examples of these spectral types are shown in Figure 2 for spectra taken on 2022 April 24 (DOY 114). The MinXSS-1 and -2 X123 sensors had smaller apertures, so their quiescent-sun (pre-flare) spectra have much less precision than the DAXSS pre-flare spectra; moreover, the DAXSS pre-flare spectra extend to much higher energy. Although Figure 2 only shows the spectra out to 4 keV for the C2 flare, the DAXSS flare spectra do extend out to about 10 keV for M-class flares as illustrated later.

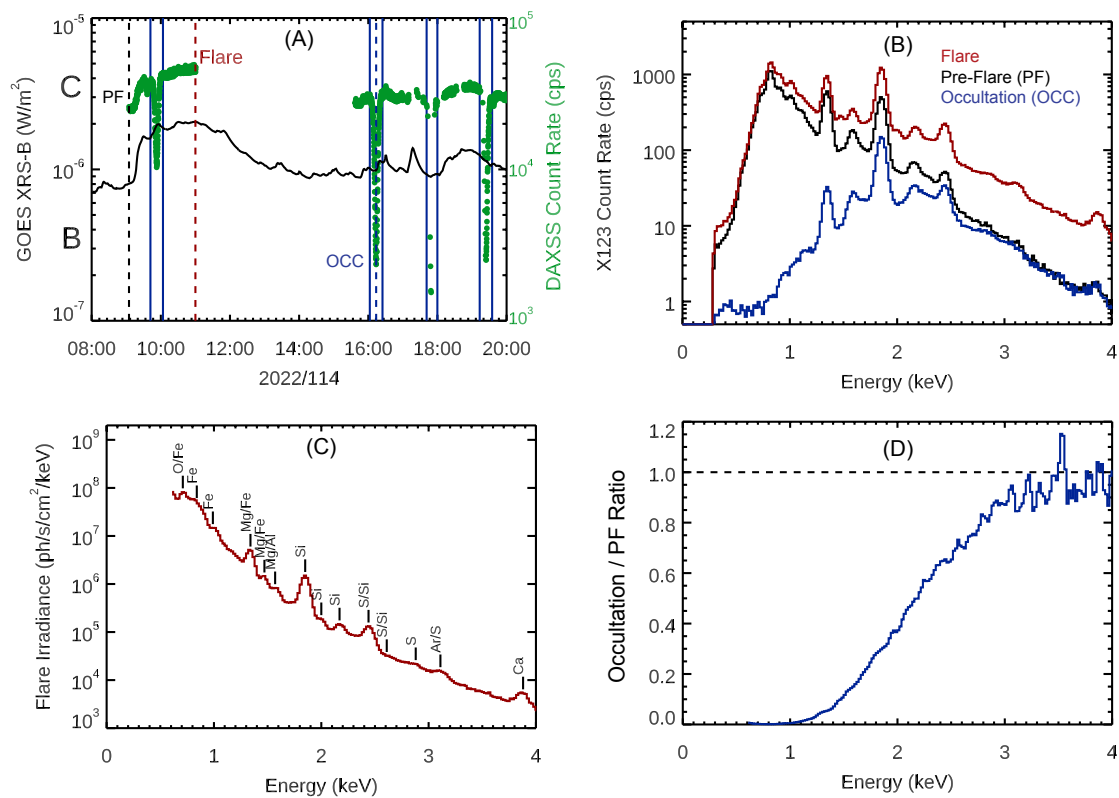

Figure 2. Example DAXSS solar spectra from 2022 April 24 (DOY 114) are for quiescent-sun (pre-flare, PF at 9:05 UT, black), flare (C2 flare at 11:00 UT, red), and occultation data (OCC at 11:14 UT, blue). The occultation data are defined as the time when the solar-viewing tangent ray height is below 300 km and are indicated as the data between the vertical blue lines in panel A. The spectra in panel B are 1-minute averages of the DAXSS Level 0D count rate spectra. The flare irradiance spectrum in panel C is from DAXSS Level 1 product with the pre-flare 9:05 UT spectrum subtracted from the flare spectrum at 11:00 UT. The solar occultation spectrum is when the tangent ray height is about 103 km and solar zenith angle is about 110°, and the ratio shown in panel D illustrates the amount of atmospheric absorption as a function of energy (wavelength).

The MinXSS-3 data are currently served from the project website at https://lasp.colorado.edu/home/minxss/ and will be archived at GSFC Space Physics Data Facility (SPDF) at the end of the mission. The current data products are at Version 2.0.0, which is the version used for all of the data analysis presented herein. In addition to the primary Level 1 science product with 9-sec cadence solar SXR spectra, there are the Level 2 and 3 products with time averages of 1-minute, 1-hour, and 1-day. The MinXSS raw packets are merged into its Level 0C and 0D data products, and those are also publicly available for additional analysis of housekeeping, SPS, and PicoSIM data.

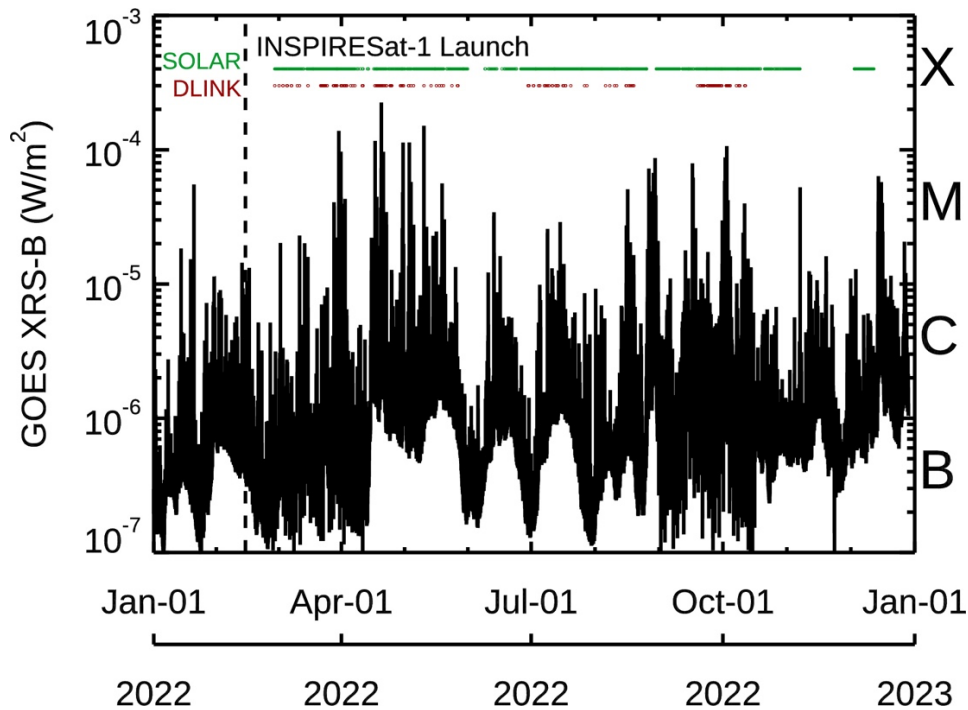

Figure 3. Solar activity over the INSPIRESat-1 mission is shown for the GOES-16 XRS-B 1-min averaged irradiance. The INSPIRESat-1 launch was on 2022 February 14. The GOES flare classes are indicated on the right axis. The times for the DAXSS solar observations are shown in green. The downlinked (DLINK) science packets as indicated in red are currently limited by the INSPIRESat-1 UHF communication system but are anticipated to be increased significantly once its S-band ground station is commissioned.

The solar activity has increased markedly since late 2021, rising from the low activity of solar cycle minimum seen in 2019-2020. The Solar Cycle 25 activity during 2022 is shown in Figure 3 for the GOES-16 XRS-B 0.1-0.8 nm irradiance. The cycle maximum is not anticipated until 2025, so the DAXSS observations in 2022 are considered to be for moderate solar activity. While INSPIRESat-1 is recording DAXSS data with 9-sec cadence and with 24-hours per day coverage for most days, the downlink of this data volume is limited, so far, to about 2% of its acquired data because only the INSPIRESat-1 UHF communication has been used for downlinks. A major thunderstorm at the India ground station site in early December 2022 caused severe damage to the ground equipment, and INSPIRESat-1 operations were limited during the 4-week repair period, but normal operations are back for INSPIRESat-1 now. Once the Indian S-band ground station is commissioned in 2023, the past DAXSS data can be downlinked over a period of several months. Fortunately, the data storage on INSPIRESat-1 is large enough on two different SD-cards that all of the DAXSS acquired data are still on-board and await downlink. Even with this limited data volume as of today, there are more than 130 DAXSS flare observations downlinked already during the INSPIRESat-1 6-month prime mission.

## 3. DAXSS Solar Occultation Experiments

For solar-observing satellites with orbit eclipse periods, they can view the atmospheric absorption profile (occultation) during the transition in and out of the orbit eclipse period. For many of these satellites, the sunset and sunrise periods are very short (few seconds) and thus don't provide enough altitude resolution for thermosphere density studies unless the solar instruments have high time cadence. For MinXSS-1 and MinXSS-2 missions, the solar occultation duration is only over one to two spectra (with 10 s cadence), and those spectra have not been studied at any detail (except to exclude them in their science data products).

The INSPIRESat-1 has a very different orbit where the view angle through the atmosphere changes slowly and solar occultation experiments can sometimes last for several minutes. Consequently, the DAXSS occultation data are intriguing because they could provide results about the thermosphere density and how much the thermosphere responds to solar SXR and EUV variations. From its SSO orbit, the occultation events are grazing slowly through the atmosphere at about 4-km per 9-sec sample when the tangent ray height is near 300 km and slows down to less than 0.5-km per 9-sec sample at the minimum tangent ray height. These altitude resolutions are much smaller than the typical 10-20 km scale heights in the thermosphere, and so analysis of the DAXSS solar occultation experiments can resolve the structure of the lower thermosphere.

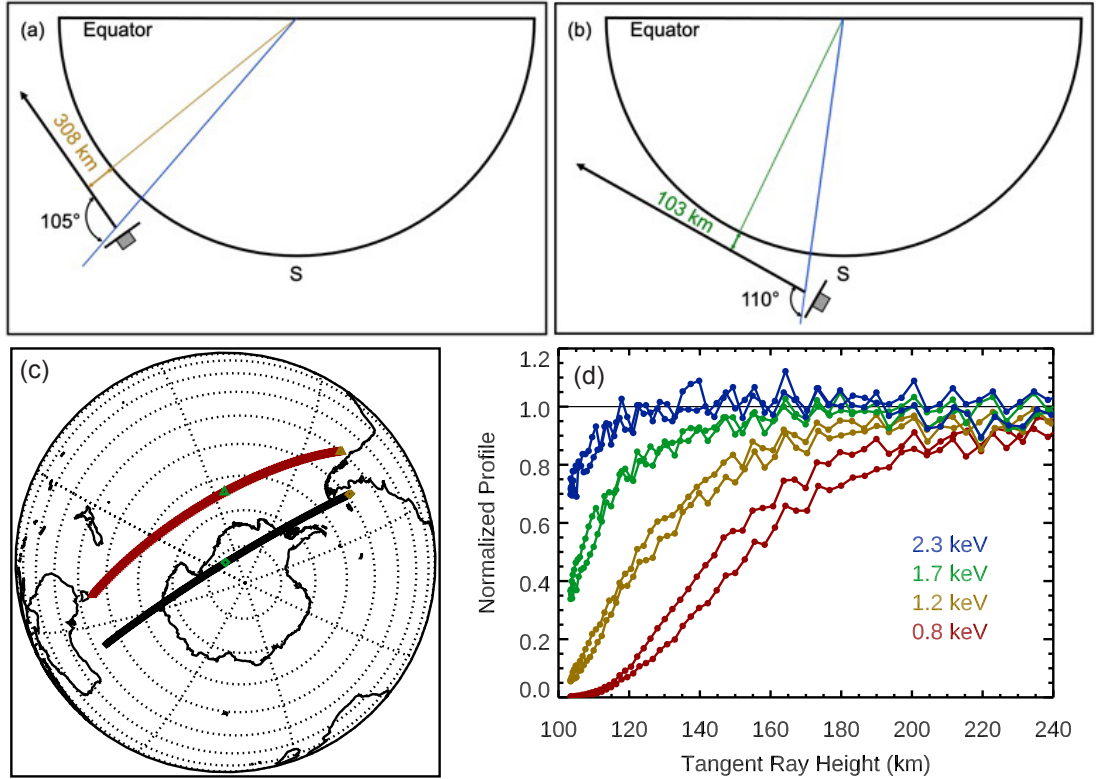

Figure 4. Example DAXSS solar occultation is illustrated for an experiment done on 2022 April 24 near 10 UT. Panels (a) and (b) show occultation geometry cartoons for the beginning of the occultation (gold) and at minimum tangent ray height (green), respectively. The occultation profiles at four different energies are shown in panel (d) using the DAXSS Level 0D data product (9-sec cadence) over a 22-min period while the spacecraft went over Earth's south pole. As illustrated in panel (c), black symbols are the spacecraft position, the red symbols are the location of the tangent ray point, the gold symbols are the starting points, and the green symbols are at the time of the minimum tangent ray height. The view of the globe is rotated so that the solar direction is straight up at the time of the minimum tangent ray height. The absorption profiles in panel (d) are normalized by the signal at the tangent ray height of 300 km and are also corrected for solar variability using the GOES XRS-B irradiance as a variability proxy. These profiles are not corrected for the different locations over Earth that changed by 169 degrees in longitude and by 37 degrees in latitude. The down-leg is the upper trace of the profiles. The spacecraft altitude and the solar zenith angle are fairly constant and with averages of 542 km and 108°, respectively.

An example of one of these occultation scans is illustrated in Figure 4. The atmospheric absorption in SXR is strongest at lower energy, and there is no appreciable absorption for altitudes above 300 km (which is the limiting altitude used for selecting solar spectra for the DAXSS Level 1 product). This absorption is consistent with photoionization of $N_2$, $O_2$, and O as their K-shell edges are near 0.5 keV. As shown in Figure 4, this DAXSS occultation is over the south pole, and the asymmetry of the down-leg and up-leg profiles is likely the result of the different legs being at different locations and different local times. Detailed thermosphere modeling is in development by co-author Robert Sewell to provide new results about the thermosphere neutral density from these solar occultation measurements in future publications.

## 4. DAXSS Solar Flare Observations

The plasma responsible for much of the solar SXR radiation is the warm (2-6 MK) corona from quiescent-sun active regions and the hot (6-16 MK) corona during flare events (e.g., Del Zanna & Mason 2018). The solar SXR spectrum includes thermal emissions from the bremsstrahlung continuum and many atomic emission lines and non-thermal emissions usually detected at the higher energies during flare events. Consequently, the analyses of the DAXSS spectra are important for studying the corona heating and cooling processes related to active region emergence and evolution and studying flare energetics. These studies can further be enhanced by combining solar SXR spectra with solar EUV observations to better understand the cool (0.5-2 MK) corona, transition region, and chromosphere (e.g., Warren 2014; Reep et al. 2020) and by combining with solar hard X-ray (HXR) observations to better understand the super-hot (16-30 MK) corona and non-thermal emissions during large flare events (e.g., McTiernan, Caspi & Warren 2019; Nagasawa et al. 2022). Here, a couple examples of DAXSS flare spectra are presented and modeling results for some of the DAXSS flare observations are presented in the next section.

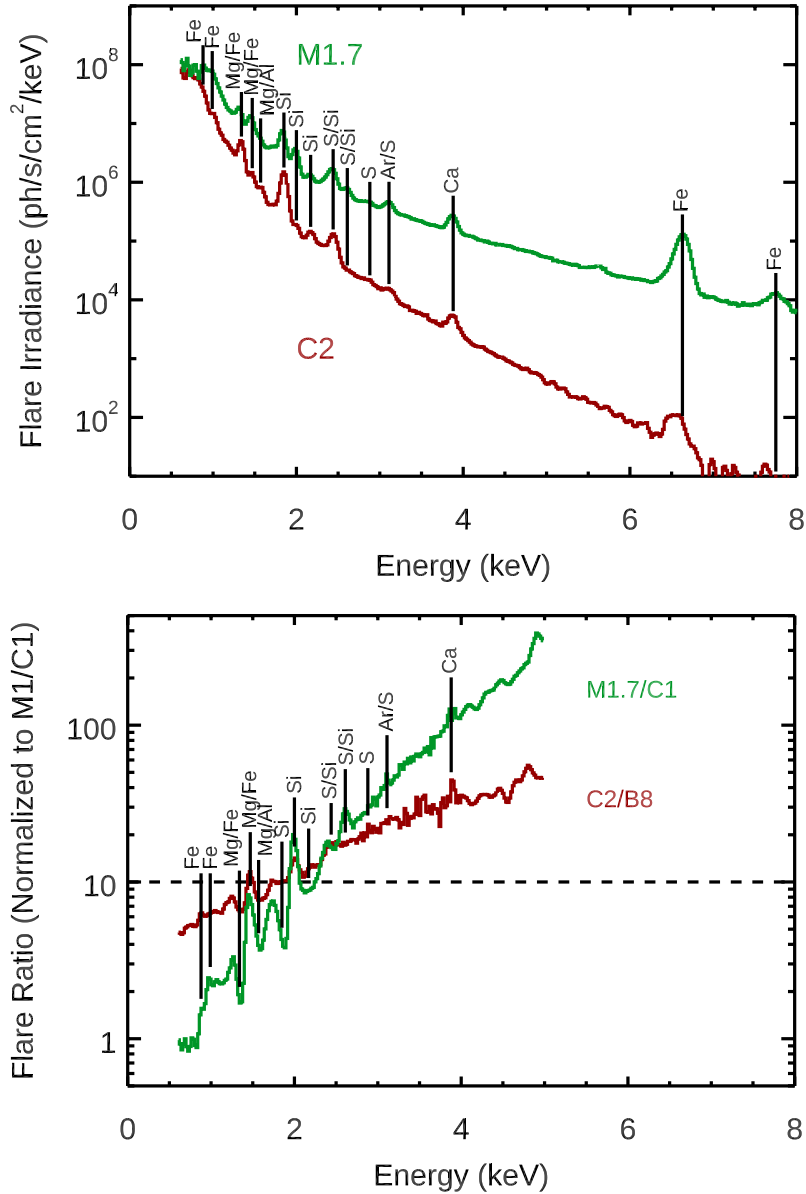

Figure 5. Example flare spectra are for the M1.7 flare (green) on 2022 March 29 (DOY 088) at 21:52 UT and for the C2 flare (red) on 2022 April 24 (DOY 114) at 11:00 UT. The top panel shows the ratio of flare peak spectrum to pre-flare spectrum and normalized for M1/C1 flare ratio (XRS factor of 10). The bottom panel shows the flare irradiance that is defined as the spectrum at the flare peak time minus the pre-flare spectrum. The ratio above 5 keV is not shown because of the low signals for the pre-flare spectra at these higher energies.

As a lesson learned from SDO EVE, we define the flare spectrum as the irradiance spectrum at the flare peak time minus the pre-flare spectrum in order to study more directly study the flaring plasma spectrum (Woods et al. 2011). Comparisons of the C2 flare on 2022 April 24 (already shown in Figure 2) and the M1.7 flare on 2022 March 29 are shown in Figure 5. We note that these flare spectra are brighter than the pre-flare irradiance at all energies, but it is interesting to note that the flare ratio (peak flare spectrum to pre-flare spectrum) is not flat and that the ratio has an upward trend towards higher energy. This upward trend with energy is the result of the flare spectrum having hotter plasma than the pre-flare spectrum. Furthermore, the steeper slope for the M1.7 flare ratio indicates that this M1.7 flare has hotter plasma than the C2 flare. This result is consistent with prior surveys that show a correlation between flare intensity and peak temperature (e.g., Caspi et al. 2014; Warmuth & Mann 2016). In addition, the relative ratio changes of different emission features for the same element at different ionization states provide another indication of the plasma temperature being different. For example, the 1.47 keV Mg/Fe feature has an enhanced flare ratio, whereas the 1.34 keV Mg/Fe feature and 1.57 keV Mg/Al feature have reduced flare ratios. This effect is due mainly to blended emission lines from different ions contributing differently for a feature as the plasma temperature changes, as discussed in more detail in Appendix A. There is also a similar behavior for the Si features, especially so for the M1.7 flare: the 2.00 keV Si feature has an enhanced flare ratio, whereas, the 1.85 keV and 2.17 keV Si features have reduced flare ratios.

5. Examples of Model Fits to DAXSS Spectra

Based on the predicted emission features as a function of plasma temperature as presented in Appendix A, we have developed some guidelines for fitting multi-temperature spectra to the DAXSS spectra. In addition to fitting plasma temperature and emission measure (EM), we are also interested in fitting abundance values to identify potential coronal heating mechanisms. Namely, the abundances of lower-energy (<10 eV) First Ionization Potential (FIP) ions, such as Ca, Fe, Mg, and Si, are typically enhanced (relative to photospheric abundances) in the quiet-Sun corona due to the ponderomotive force related to Alfvén-waves (e.g., FIP-effects are described by Laming 2015). This enhancement often appears to be reduced during flares (e.g., Warren 2014). For example, the low-FIP ions can have enhanced coronal emissions in quiescent-Sun spectra relative to high-FIP ion emissions because the corona abundances can be enhanced for the low-FIP ions, and then those low-FIP corona abundances can decrease during flares as related to injection of heated plasma into the corona from the chromospheric evaporation process. Ideally, one wants to fit abundances for all of the emission features and with as many temperatures as needed to improve the spectral fit to the DAXSS spectra. But as shown from the analysis of the SXR spectral features in Appendix A, there are limitations with current techniques for which elemental abundances can be fit as summarized in Table 1 and described next.

Table 1. Abundance Modeling Guidelines for DAXSS Spectra
"Fixed" means to use the Feldman (1992) extended coronal abundances.
"Vary" means to allow that element's abundance to be a free parameter.

| Modeling Parameter | < 4 MK | 4–8 MK | > 6 MK |
|---|---|---|---|
| Solar Component | Quiet-Sun | Active Region | Flare |
| DAXSS Energy Range | 0.6–3 keV | 0.6–4 keV | 0.6–7 keV |
| Ar Abundance | Fixed | Vary | Vary |
| Ca Abundance | Fixed | Vary | Vary |
| Fe Abundance | Fixed | Fixed/Vary | Vary |
| Mg Abundance | Fixed/Vary | Vary | Vary |
| Si Abundance | Fixed/Vary | Vary | Vary |
| S Abundance | Fixed | Vary | Vary |
| Other Elements | Fixed | Fixed | Fixed |

To illustrate these fitting techniques for DAXSS spectra, we provide example fits for the multiple C-class flares on 2022 April 24 (see Figure 2A time series plot). The example of model fitting shown in Figure 6 uses a two-temperature (2T), two-emission-measure (2EM) fit with separate abundance factors (AFs) for Mg, Si, S, and Fe. These AFs, also called the FIP bias, are relative to photospheric abundance values; i.e., a value of 1.0 indicates a photospheric abundance. For example, the ratios of the Feldman (1992) extended coronal abundances to the Asplund et al. (2021) photospheric abundances are 4.36 for Fe and Ni, 3.98 for Mg, 3.89 for Si, 3.27 for Ca, 1.58 for Ar and O, 1.41 for S, and 1.05 for Ne. The AF values for the low-FIP ions (Al, Ca, Mg, Ni, Fe, Si) are expected to be about four for quiet-Sun (no active regions) (e.g., Feldman 1992), about two for quiescent (non-flaring) active regions (e.g., Del Zanna & Mason 2018), and as low as one during large flare events (e.g., Warren 2014). These are our general expectations for the abundance factors with solar activity for the solar SXR spectra, but there are conflicting results based on solar EUV spectra. One example is the photospheric abundances for coronal quiet-Sun and active-region emissions in the EUV range as reported by Del Zanna (2019). Those results from EUV spectra are for low-FIP emissions at temperatures less than 1.5 MK; whereas, the solar SXR spectra tend to be dominated with emissions at temperatures more than 2 MK. As our following analyses are only with SXR spectra, those differences in abundances for EUV and SXR spectra will not be addressed here.

This fitting process uses existing IDL SolarSoftWare (SSW; Freeland & Handy 1998) code with the atomic database from CHIANTI Version 10 (Dere et al. 1997, 2019) with ionization fractions from Mazzotta et al. (1998) to generate solar spectra that contains both the emission lines and the underlying thermal background continuum within the selected energy ranges. The primary IDL function is f_vth.pro, written by Richard Schwartz (Schwartz et al. 2002). It has the advantage of interpolating between a precomputed sample space of CHIANTI solutions with various temperatures and emission measures, which saves a large amount of computation time. The f_vth.pro function then outputs a model solar spectrum for a given energy range when provided a temperature, emission measure, and relative elemental abundance factor as input parameters. The relative elemental abundance factor used is a scaling factor that is multiplied by a standard (or custom) list of elemental abundances. For this model the Feldman Standard Extended Coronal elemental abundance values (Feldman et al. 1992; Landi et al. 2002) were used as the reference coronal abundances. These abundances are held fixed for the lower temperature component in the model and allowed to vary for low-FIP ions (Ca, Fe, Mg, and Si) for the higher temperature component. Changes in abundance values for this spectral model affect both emission lines and continua. We also note that the model spectra are dependent on the choice of which abundance set is used in CHIANTI for the elements whose abundance values are fixed.

Each minute-averaged DAXSS spectrum is fit independently using this modeling method and the modeled spectra are compared to the measured spectra to check their goodness of fit. The input parameters for each of the 2T, 2EM, and 4AF are iteratively adjusted until a least-squares minimum $\chi^2$ value is found. This is done using the IDL procedure mpfit.pro, which uses a Levenberg–Marquardt technique to iteratively find the best-fit parameter values.

For DAXSS quiescent-sun (pre-flare) spectra, two-temperature (2T) fits are needed (e.g., Schwab et al. 2020), and fitting with abundances fixed to the Feldman (1992) extended coronal abundances (FSEC) usually provides good fit results. For improved DAXSS spectral fits, especially so if active regions are present, one can fit with a quiet-sun component near 1-2 MK with fixed abundances at the Feldman coronal values and a warmer quiescent-active-region component above 3 MK with an adjustable abundance for the low-FIP elements. When fitting the abundances for quiescent-sun DAXSS spectra for this warmer component, we recommend just fitting the Mg, Si, and S abundances with the spectral range below 3 keV. Fitting the abundances for Ca, Fe, Ne, and O for the quiescent-sun DAXSS spectra has not provided as accurate results, probably because of the changing contributions with temperature of those ions to the low-energy features near 1 keV and the significant blending within those features (see Figure A.1 and discussion in Appendix A). The low-FIP elements (Mg, Si, Fe) have FIP bias between two and three for 2022 April 24, and the mid-FIP S element has a FIP bias near one. The review of spectral modeling by Del Zanna & Mason (2018) provides several abundance results indicating a lower FIP bias for active regions, and their results are similar for the pre-flare model fits shown in Figure 6.

For DAXSS flare spectra, three-temperature (3T) and four-temperature (4T) models can provide good fit results with the two lower temperatures (often near 2 MK and 4 MK) representing the quiescent-sun portion of the DAXSS irradiance spectra and the higher temperatures (usually between 6 MK and 16 MK) representing the flaring plasma. For fitting flare abundance changes, we usually fit the lower temperature component with fixed Feldman (1992) coronal abundances and allow the fit to adjust the abundance of Ca, Fe, Mg, Si, and S for the higher temperature component. The flare Fe abundance fits are more accurate if the DAXSS spectra extends above 6.8 keV to include the high-temperature, highly-ionized-Fe feature near 6.6 keV; otherwise, the lower-energy Fe emissions have competing contributions from blended ions (see Appendix A) and also from the quiescent-sun and flare components.

For these smaller C-class flares on 2022 April 24, it was found that only two-temperature fitting method was needed to obtain good fits. The two temperature results from these model fits are shown in Figure 6 and are also compared with the single-temperature and single-emission-measure results found using the GOES-16 XRS A/B ratio method, whereby the flare temperature ($T_{XRS}$) is estimated by Equation 1 and the flare emission measure ($EM_{XRS}$) is estimated by Equations 2 and 3.

$$T_{XRS}(MK) = 2.7460 + 129.47 \cdot R - 966.28 \cdot R^2 + 5517.5 \cdot R^3 - 1.8664 \times 10^4 \cdot R^4 \quad (1)$$
$$+ 3.5951 \times 10^4 \cdot R^5 - 3.6099 \times 10^4 \cdot R^6 + 1.4687 \times 10^4 \cdot R^7$$

$$B_{model} = 6.9469 - 6.0827 \cdot T_{XRS} + 1.7364 \cdot T_{XRS}^2 - 0.15594 \cdot T_{XRS}^3 \quad (2)$$
$$+ 6.7848 \times 10^{-3} \cdot T_{XRS}^4 - 1.4446 \times 10^{-4} \cdot T_{XRS}^5 + 1.2089 \times 10^{-6} \cdot T_{XRS}^6$$

$$EM_{XRS}(10^{55} \text{ cm}^{-3}) = B_{measure}/B_{model} \qquad (3)$$

The R variable is the ratio of the XRS-A irradiance (A) to XRS-B irradiance (B) as reported by NOAA for GOES-16. The A and B variables are irradiances in units of W m$^{-2}$ for the XRS-A and XRS-B channels, respectively. This relationship to temperature was determined by Tom Eden by convolving the XRS responsivities with isothermal CHIANTI model solar spectra using Feldman et al. (1992) extended coronal abundances over the range from 3 MK to 32 MK and then fitting a seventh-order polynomial. This XRS A/B ratio method for temperature is very similar to that of Garcia (1994) and White, Thomas, & Schwartz (2005), but the relationships are different for the GOES-R series because the detector technology changed to Si photodiodes for the GOES-R series XRS. This estimated XRS temperature, in units of MK, is used in Equation 2 to calculate the model XRS-B normalized response ($B_{model}$), and then Equation 3 is used to calculate the XRS emission measure with this $B_{model}$ value and XRS-B irradiance, $B_{measure}$ (see White, Thomas, & Schwartz (2005) for more details on this method). We note that double precision calculations are required to use these high-order polynomial equations. The root-mean-square residuals for these XRS model fits are 0.08 MK for XRS temperature ($T_{XRS}$) and 0.03x10$^{55}$ cm$^{-3}$ for the XRS-B normalized response ($B_{model}$). However, the assumption for the elemental abundance values has larger uncertainties for this simple XRS model. S. White (priv. comm.) did a similar analysis with GOES XRS data using both coronal and photospheric abundances and finds that the estimated XRS temperature is within 1 MK using either abundance set, but finds that the estimated EM with photospheric abundances is about a factor of two larger than the EM with coronal abundances.

As seen in Figure 6, the XRS single-temperature estimate (T_XRS) is slightly higher than the warmer temperature (T2) from our 2T model. Both the T_XRS and T2 increase and decrease with similar behavior as the GOES-16 XRS-B irradiance trend. There are two main reasons for the XRS single-temperature estimate being higher than our 2T model result. The XRS temperature model (Equation 1) is for coronal abundance, but as the abundances decrease during a flare, the XRS temperature model with photospheric abundances would predict different temperatures (White, Thomas, & Schwartz 2005). In particular, if the true abundances are lower than assumed, the model would require a higher T and/or higher EM to account for the same irradiance level, as seen here. The other reason is because the GOES XRS passbands are at higher energy (1.5-25 keV) than the DAXSS spectral range (0.6-9 keV) and consequently the GOES XRS measurements are both more sensitive, and more weighted towards, hotter plasma. This comparison provides validation that both methods appear to provide reasonably accurate results as both have very similar trends. Overall, these models for the hotter temperature component are in agreement with each other to within ~1 MK.

The XRS single-emission-measure estimate (EM_XRS) has similar trend to the warmer emission measurement (EM2), but it is scaled down by factor of 0.25 in Figure 6 top panel for this comparison. The primary reason for needing this scaling for the XRS EM estimate is because the XRS model is for coronal abundance, which is about a factor of four higher for the low-FIP elements. Another reason for the scale difference is that the T_XRS is higher, which thus generally cause lower EM values when modeling the corona.

The GOES flare class of the larger flare is about C5. For even larger (more intense) flares there are higher photon fluxes that provide an increased number of photon counts across the entire DAXSS energy range. This provides spectra with better counting statistics, that is, higher signal-to-noise ratio. Particularly for the elements with emission features at the higher energies such as S, Ca, and Ar, modeling the DAXSS spectra for the AFs of these elements is only feasible for the larger M-class flares. For this C5 flare and using 1-minute integrations, the AF for S is able to be fit with confidence, but with more intense flares of M1 or higher, it is possible to include AF fitting of Ca, Ar, and even the high energy Fe features around 6-8 keV in the model. A higher signal-to-noise ratio may also be achieved using longer integration times, such as an hour in some cases. This is particularly helpful for providing abundance calculations for when solar conditions are reasonably consistent over the integration period, such as quiescent periods between flare times, but would not be applicable for flaring times when conditions change on timescales of minutes or faster.

The evolution of the hotter temperature component and magnitude of the low-FIP ion abundances during flares, as well as the phasing of those changes, are important indicators for the coronal heating processes. The sudden decreases of the low-FIP ion abundances during the flare impulsive phase, that is, prior to the gradual phase rise of the XRS irradiance, is very supportive of the chromospheric evaporation process being a key part of the flare irradiance variation, as also noted by Warren (2014). The abundance magnitudes decrease to near photospheric levels, as well as the amount of the rise of the plasma temperature, are indicators of the amount of the energy released during the flare, presumably from the initial magnetic reconnection in the coronal loops.

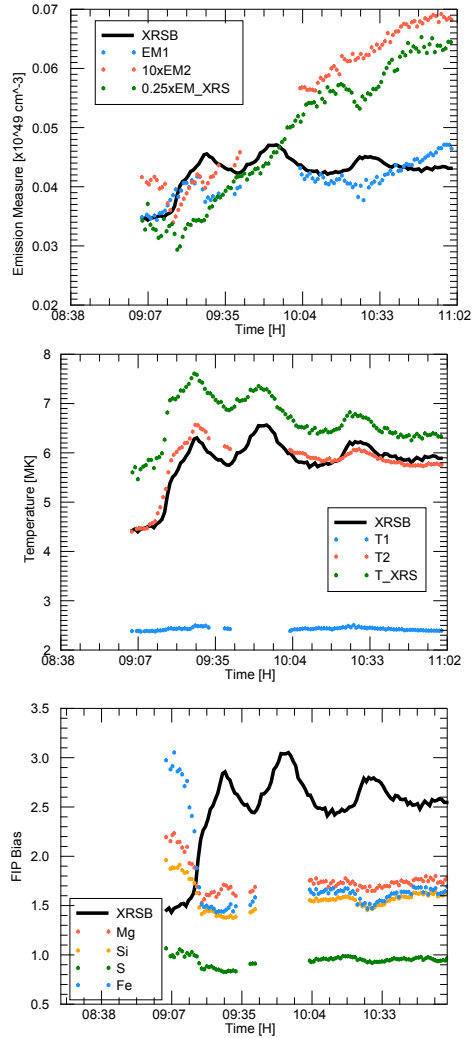

Figure 6. Analysis of emission measure (EM), temperature (T), and abundance factor (AF) evolution spanning the multiple flares on 2022 April 24. Two EMs are shown in the top plot and their respective two Ts are shown in the middle plot. For all three plots, a scaled GOES-16 XRS-B irradiance (black line) is shown to indicate the evolution of the flare. The cooler temperature component, T1, remains relatively stable across the duration of the flares, whereas the warmer temperature component, T2, varies along with the scaled XRS-B irradiance. These temperatures are compared with the single-temperature result found using the A/B ratio method from GOES-16 (T_XRS). The XRS single-temperature estimate is slightly higher than the warmer temperature in our 2T model. The abundance factors (AF, FIP bias) of Mg, Si, S, and Fe are shown in the bottom plot to suddenly decrease from their pre-flare values at the beginning of the flare and slightly decrease during the rise of each sequential flare impulse and then relax slowly toward pre-flare values between impulses.

Further insight into how the AF changes with temperature over the course of this flare can be seen in Figure 7. From this figure an semi-linear relationship between AF and T emerges, showing that as the temperature during the flare increases there is a decrease in AF for the low-FIP elements Mg, Si, and Fe as well as smaller decrease for the mid-FIP element S. Further analysis into this relationship with many more flare events is needed to determine if this physical phenomenon is real or a fitting artifact resulting from using this simplistic 2T model for a limited data sample.

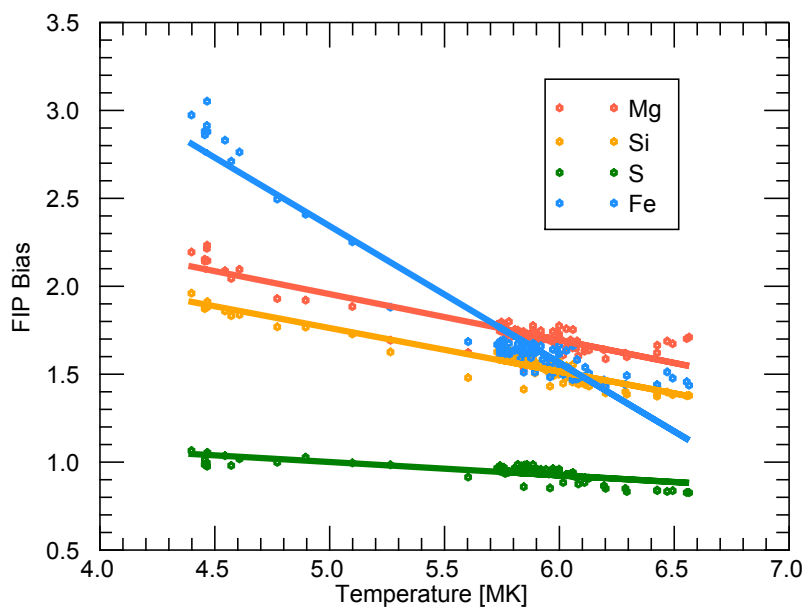

Figure 7. The abundance factors (AF) of Mg, Si, S, and Fe, as plotted against the hotter temperature component (T2) of the 2T model fit, show a semi-linear relationship. The AFs of these low- and mid-FIP elements decline as the flare temperature increases.

Spectral modeling of solar SXR spectra from other instruments have yielded similar results. For example, Moore et al. (2018) show 2T model results for MinXSS-1 pre-flare spectra having temperatures in the 1.2 MK to 5.2 MK range and flare spectra having a hotter component in the 13 MK to 20 MK range. Moore et al. (2018) don't provide time series for abundance changes during flares, but they do find significant reduction of Fe, Mg, Si, and Ni abundances for three flare spectra relative to their pre-flare abundances. The analyses of abundance values during 526 flares observed by MESSENGER SAX indicate that the Fe, S, and Si abundances are near photospheric levels but that the Ca and Ar abundances are closer to the enhanced coronal abundance levels (Dennis et al. 2015). Narendranath et al. (2020) also present similar results for abundance changes during several flares using 2-20 keV spectra from MESSENGER SAX and SMART-1 XSM.

There are also modeling results for dozens of A-class and B-class flares observed by Chandrayaan-2 XSM. Those XSM results from 1T modeling indicate flare temperatures in the 4 MK to 8 MK range and near photospheric abundances for Al, Mg, Si, and S during flares (Nama et al. 2023; Mondal et al. 2021). For example, Mondal et al. (2021) found decreases in the abundance of low-FIP ions during flares using Chandrayaan-2 XSM spectra in the 1-15 keV range. They show for nine B-class flares that the abundances for Al, Mg, Si, and S start off near enhanced-coronal abundance levels before the flare and that those elemental abundances decrease to near photospheric levels near the SXR peak and then increase slowly back towards coronal levels. We note that Mondal et al. (2021) also provide two scenarios for explaining the low-FIP abundance changes during flares and describe in detail their XSPEC modeling code. Nama et al. (2023) also show similar results for many A-class flares using XSM spectra, and they note that the flare-related abundance decreases for Al, Mg, and Si appear to be proportional to the flare magnitude, and is somewhat similar to what is shown in Figure 7 for the DAXSS abundance variations.

Similarly, Sylwester et al. (1984, 2022, 2023; and references therein) showed evidence for decreasing Ca abundance in a significant number of flares observed with the Bent Crystal Spectrometer on the Solar Maximum Mission. In addition, X-ray measurements with the RESIK crystal spectrometer showed potential decreases in Si abundance (Sylwester et al. 2014).

All of these results – from DAXSS, XSM, SMM, RESIK, and others – highlight the importance of chromospheric evaporation during flares to inject warmer plasma into the coronal loops and is even suggestive that the amount of chromospheric evaporation is related to flare temperature and intensity.

## 6. Conclusions and Future Work

The higher-resolution SXR spectra from DAXSS are providing more spectral features than those from MinXSS-1 or -2 instruments and thus providing new information for exploring the highly variable corona. Modeling the multiple flares on 2022 April 24 illustrates the capability to fit DAXSS spectra with multiple temperatures and estimating the abundances of Fe, Mg, Si, and S. These model results indicate (1) quiescent-sun (pre-flare) contributions in the 2-3 MK range, (2) flare temperatures in the 6-7 MK range for the ~C5 flare, and (3) reduction of low-FIP ion abundances during flares for Fe, Mg, Si, and S. Similar results have been found for brighter flares, as expected, but with hotter plasma temperatures in the 10-15 MK range. The phasing of the abundance reductions during the flare impulsive phase is strong support for the chromospheric evaporation process during flares.

Comparisons to other SXR and HXR data sets (e.g., Hinode Solar X-Ray Telescope, Chandrayaan-2 XSM, and Solar Orbiter STIX) and analyses of many more flare events are in progress. These additional analyses are anticipated to reveal new results about flare onset energetics, solar conditions for the hotter components in the SXR spectra, and phasing of the elemental abundance changes during active region evolution and during flares. Comparisons with measurements more sensitive to hotter components, such as measured by RHESSI and/or STIX, are particularly important to help understand whether chromospheric evaporation is a strong contributor to those hotter components, or whether direct-heating processes are dominant (e.g., Longcope & Guidoni 2011; Caspi et al. 2015).

These new SXR spectra are also important for updating solar irradiance spectral models (e.g., FISM-2, Chamberlin et al. 2020), which in turn will be useful to improve understanding of the ionosphere formation and thermosphere heating for terrestrial and planetary atmosphere/ionosphere studies. In addition, the DAXSS occultation measurements are being studied to examine thermosphere density changes with different solar activity levels.

In the longer-term future, these DAXSS results highlight the need for systematic observations of SXR spectra from flares and quiescent active regions, with improved spectral resolution to mitigate spectral-line blends and provide better accuracy for fitting of temperatures and abundance factors. The Marshall Grazing-Incidence X-ray Spectrometer (MaGIXS, Savage et al. 2023) sounding rocket demonstrated a novel slitless spectrograph for providing spatially and spectrally resolved SXR measurements. Although their first flight studied an X-ray bright point, future flights and/or a future space-borne instrument could provide excellent abundance measurements of active regions and flares. Similarly, the CubeSat Imaging X-ray Solar Spectrometer (CubIXSS, Caspi et al. 2022), leveraging the pathfinding MinXSS and DAXSS developments and measurements, will fly multiple Amptek X123 FastSDDs along with a different novel slitless spectrograph to provide high-resolution SXR spectra over an unprecedented range from 1 to >55 Å. With an expected launch in mid-2025, CubIXSS will study flares and active regions during the peak of Solar Cycle 25, including time-resolved studies of flare heating and active region evolution. These DAXSS measurements will guide future studies by these missions to better understand flare and active region heating processes.

## 7. Acknowledgments

NASA grant NNX17A171G has supported the flight build, testing, and operations for the DAXSS instrument. We thank the many staff and students at University of Colorado in Boulder, Indian Institute of Space Science and Technology in India, Nanyang Technological University in Singapore, and National Central University in Taiwan for the development, testing, and operations of the INSPIRESat-1 small satellite. We are also very grateful to the Indian Space Research Organization for providing the launch. AC was also partially supported by funding from NASA grant 80NSSC19K0287 and NASA cooperative agreement 80NSSC22M0111. We thank Peter Young for making updates for CHIANTI Version 10.2 to include annotation for the dielectronic recombination emission lines in the CHIANTI line list files; this update to CHIANTI was critical for having ion levels more correctly listed in our Appendix A tables. We also thank Iain Hannah for sharing his CHIANTI Version 10 files for use with the IDL f_vth.pro modeling of the DAXSS spectra and thank Stephen White for sharing his information about SXR spectral modeling with different abundance values.

## 8. Data Source

The DAXSS solar SXR spectral irradiance data products, along with user guide and data plotting examples in IDL and Python, are available from the MinXSS web site at http://lasp.colorado.edu/home/minxss/. The GOES XRS data are available from http://www.ngdc.noaa.gov/stp/satellite/goes/dataaccess.html.

## 9. References

Aschwanden, M. J. 2019, New Millennium Solar Physics - Astrophys. Space Sci. Lib. 458 (Cham: Springer Nature Switzerland AG)

Asplund, M., Amarsi, A. M., & Grevesse, N. 2021, A&A, 653, A141

Caspi, A., Barthelemy, M., Bussy-Virat, C. D., et al. 2022, Space Weather, 20, e2020SW002554

Caspi, A., Krucker, S., & Lin, R. P. 2014, ApJ, 781, 43

Caspi, A., Shih, A. Y., McTiernan, J. M., & Krucker, S. 2015, ApJL, 811, L1

Chamberlin, P. C., Eparvier, F. G., Knoer, V., et al. 2020, Sp. Weather, 18, 12

Chandran, A., Fang T.-W., Chang, L., et al. 2021, Adv. Space Res., 68, 2616

Del Zanna, G., 2019, A&A, 624, A36

Del Zanna, G., Mason,H. E. 2018, Living Rev. Sol. Phys., 15, 5

Del Zanna, G., Dere, K. P., Young, P. R., and Landi., E. 2021, ApJ, 909, 38

Dennis, B. R., Phillips, K. J. H., Schwartz, R. A., et al. 2015, ApJ, 803, 67

Dere, K. P., Landi, E., Mason, H. E., Monsignori Fossi, B. C., & Young P. R. 1997, A&A, 125, 149

Feldman U., Mandelbaum P., Seely J. F., Doschek G. A. and Gursky H. 1992 ApJS 81 387

Freeland, S. L., & Handy, B. N. 1998, Sol. Phys., 182, 497

Garcia, H. A. 1994, Sol. Phys., 154, 275

Gburek, S., Sylwester, J., Kowalinski, M., et al. 2011, SoSyR, 45, 189

Krucker, S., Hurford, G. J., Grimm, O., et al. 2020, A&A, 642, A15

Laming, J. M. 2015, Living Rev. Sol. Phys., 12, 2

Lin, R. P., Dennis, B. R., Hurford, G. J., et al. 2002. Sol. Phys., 210, 3

Longcope, D. W., & Guidoni, S. E. 2011, ApJ, 740, 73

Mason, J. P., Woods, T. N., Caspi, A., et al. 2016, J. Spacecraft Rockets, 53, 328

Mason, J. P., Baumgart, M., Rogler, B., et al. 2017, J. Small Sat., 6, 651

Mason, J. P., Woods, T. N., Chamberlin, P. C., et al. 2020, Adv. Space Res., 66, 3

Mazzotta P., Mazzitelli G., Colafrancesco S. and Vittorio N. 1998 A&AS 133 403

McTiernan, J. M., Capsi, A., & Warren, H. P. 2019, ApJ, 881, 161.

Mithun, N. P. S., Vadawale, S. V., Sarkar, A., et al. 2020, Sol. Phys., 295, 139

Mondal, B., Sarkar, A., Vadawale, S. V. et al. 2021, ApJ, 920, 4

Moore, C. S., Woods, T. N., Caspi, A., & Mason, J. P. 2016, Proc. SPIE 9905, 990507

Moore, C. S., Caspi, A., Woods, T. N., et al. 2018, Sol. Phys., 293, 21

Nama, L., Mondal, B., Narendranath, S., Paul, K. T. 2023, Sol. Phys., 298, 55

Nagasawa, S, Kawate, T., Narukage, N. et al. 2022, ApJ, 933, 173

Narendranath, S., Sreekumar, P., Pillai, N. S. et al. 2020, Sol. Phys., 295, 175

Pesnell, W. D., Thompson, B. J., & Chamberlin, P. C. 2012, Sol. Phys., 275, 3

Phillips, K. J. H. 2004, ApJ, 605, 921

Reep, J. W., Warren, H. P., Moore, C. S., et al. 2020, ApJ, 895, 30

Savage, S., Winebarger, A., Kobayashi, K., et al. 2023, ApJ, 945, 105

Schlemm, C. E., Starr, R. D., Ho, G. C., et al. 2007 Space Sci. Rev., 131, 393

Schwab, B. D., Sewell, R. H. A., Woods, T. N., et al. 2020, ApJ, 904 20

Schwab, B. D., Woods, T. N., & Mason, J. P. 2023, ApJ, 945, 31

Schwartz R. A., Csillaghy A., Tolbert A. K. et al. 2002, SoPh, 210, 165

Spence, H. E., Caspi, A., Bahcivan, H., et al. 2022, Space Weather, 20, e2021SW003031

Sylwester, J., Kowalinski, M., Gburek, S., et al. 2012, ApJ, 751, 111

Sylwester, J., Lemen, J. R., & Mewe, R. 1984, Nature, 310, 665

Sylwester, B., Sylwester, J., Phillips, K. J. H., Kępa, A., & Mrozek, T. 2014, ApJ, 787, 122

Sylwester, J., Sylwester, B., Phillips, K. J. H., & Kępa, A. 2022, ApJ, 930, 77

Sylwester, B., Sylwester, J., Phillips, K. J. H., & Kępa, A. 2023, ApJ, 946, 49

Ward, W., Seppälä, A., Yigit, E. et al. 2021, Prog. Earth Planet. Sci., 8, 47

Warmuth, A., & Mann, G. 2016, A&A, 588, A115

Warren, H. P. 2014, ApJLett, 786, L2

White, S. M., Thomas, R. J., Schwartz, R. A. 2005, Sol. Phys., 227, 231

Woods, T. N. Elliott, J. 2022, Sol. Phys., 297, 64

Woods, T. N., Hock, R., Eparvier, F., et al. 2011, ApJ, 739, 59

Woods, T. N., Caspi, A., Chamberlin, P., et al. 2018, ApJ, 835, 122

Appendix A. Solar Spectral Features in the Soft X-ray (SXR) Range

To properly interpret the solar spectral variations and the modeling results concerning elemental abundances, it is important to understand the solar emission features, such as the various contributions from multiple line blends at the instrument spectral resolution and how those contributions change with plasma temperature and density. There are 17 distinct spectral features identified in the DAXSS flare spectra as shown in Figures 2 and 5, and modeling the SXR spectra at the DAXSS resolution suggests that 19 different spectral features can be prominent depending on the plasma temperature. This appendix describes the results of modeling the SXR spectra at the DAXSS energy resolution.

Multiple-temperature fits of the DAXSS and MinXSS quiescent-sun spectra usually indicate coronal temperatures in the 2 MK to 8 MK range (e.g., Schwab, Woods, & Mason 2023; Moore et al. 2018), and fits to flare spectra usually have an additional hotter component in the 8 MK to 16 MK range (e.g., Moore et al. 2018) and sometimes a super-hot component near 30 MK (Nagasawa et al. 2022). Of course, the coronal plasma temperatures are not discrete and are distributed continuously over temperature (e.g., see Differential Emission Measure (DEM) fits for MinXSS spectra from Reep et al. 2020; Woods & Elliott 2022). Nonetheless, it is useful to examine the emission feature contributions for the iso-thermal plasma as listed in Tables A1-A6. The emission lines are estimated using the CHIANTI atomic database for isothermal calculation with Feldman (1992) extended coronal abundances (CHIANTI version 10.2 code, version 10.1 database, Dere et al. 1997; Del Zanna et al. 2021). The DAXSS energy resolution (Schwab et al. 2020) is applied first to the CHIANTI spectra and then the peaks are identified. The many emission lines near each peak are weighted by a Gaussian function to approximate the DAXSS energy resolution and then a smoothed spectrum is used to get center-of-energy value and percent contribution for all ions. Only the top three contributing ions are listed in Tables A.1-A.6 for each feature.

The CHIANTI version 10.2 code now provides an "s" annotation for ion levels in their standard line-list output files to denote emissions from transitions due to inner shell excitation and dielectronic recombination. Earlier versions of CHIANTI do not provide this annotation. For emission lines with the "s" annotation, we assigned those lines as belonging to the next-higher ion level for the Appendix A table listing. For example, an Fe XXIVs line is assigned to be an emission from Fe XXV. This is appropriate for dielectronic recombination emission lines because the "parent" ion for the recombination actually is one level higher (e.g., Fe XXV). While inner shell excitation is indeed from the lower-level ion (e.g., Fe XXIV), those transitions require higher energies and thus occur at a similar temperature as the dielectronic recombination lines (P. Young 2023, priv. comm.); therefore, we treated these transitions the same way, incrementing the ion level for all lines annotated with an "s" as representative of a hotter ion. In other words, the ion level percentages as listed in the Appendix A tables have some additional uncertainty due to the interpretation of the ion level "s" annotations. We note that the element percentages are more accurate as they are not affected by our interpretation of the ion level "s" annotations.

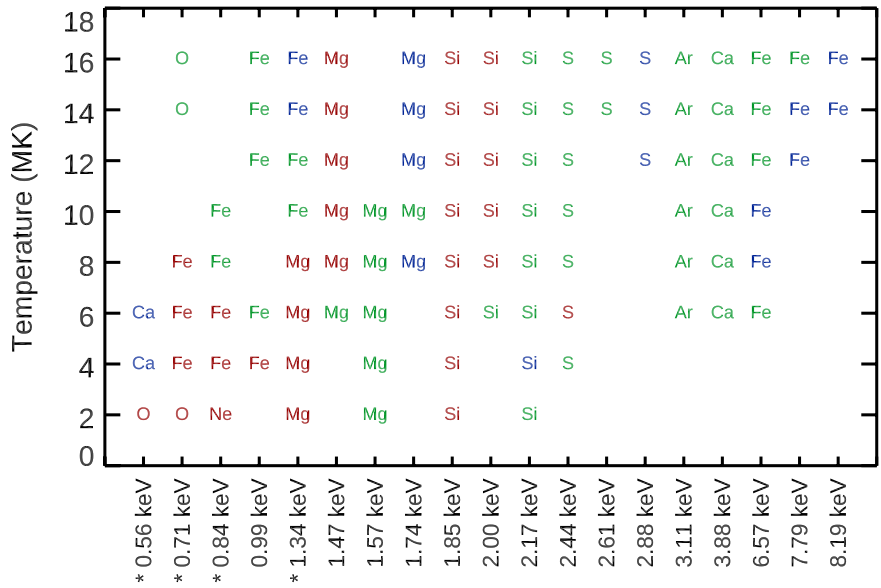

Figure A.1. The features expected in the DAXSS spectra are illustrated as a function of plasma temperature and are labeled with the primary ion as listed in Tables A.1-A.6. The color of the labels is related to the intensity ratio relative to the nearby continuum intensity; features with a ratio greater than 2.0 are red, with ratio between 1.3 and 2.0 are green, and with ratio less than 1.3 are blue. An asterisk by the feature energy label signifies that the primary ion changes with plasma temperature.

Not all features listed in Tables A.1-A.6, and illustrated in Figure A.1, are identified for all temperatures in the 2–16 MK range. For example, the higher energy features (Ar, Ca, and Fe in the 3-8 keV range) are only measured in the flare spectra, and those features start to show up at 6 MK and hotter. Depending on temperature, the accuracy for ion abundances is better over a more limited temperature range for some features. For example, the lower energy features between 0.5 keV and 1.4 keV have competing ion dominance between Ca, Fe, Mg, Ne, and O; consequently, abundance results from using just the lower-energy range tend to be less accurate, and higher-energy features are needed to better constrain the competing elements. The most prevalent ions over the 2–16 MK range are the Mg, Si, and S features between 1.4 keV and 2.9 keV, although the S features are not bright at 2 MK and a couple of S features only appear at the hotter temperatures. In other words, the DAXSS spectra can provide Mg, Si, and S abundance results for both quiescent-sun (lower temperature) spectra and flare (higher temperature) spectra, and DAXSS flare spectra can provide additional information on abundances for Ar, Ca, and Fe. Notably, it is a challenge to get accurate Fe abundance for temperatures below 6 MK for SXR spectra that are limited in energy range below 5 keV due to the blending of other ions in the Fe-related features near 1 keV.

In addition to how the primary ion percent contribution can change with temperature, the features listed in Tables A.1-A.6 also illustrate how the energy of some features can peak at different energies as the temperature changes. For example, the Fe feature near 6.5 keV changes from a peak energy of 6.42 keV at 6 MK to 6.65 keV at 16 MK; this being a consequence of higher levels of Fe ionization with hotter temperatures affecting the distribution of the many different Fe emissions near 6.5 keV (see Phillips 2004).

Some of these results about solar SXR emission features are discussed more in the above sections about flare variability and modeling of elemental abundances. The following describes the various emission contributions to each of the spectral features. The following Appendix A tables list the features' peak energy, ratios of features' intensity peak to continuum, percent contribution for the three dominant elements, and percent contribution for the three dominant ions. These tables list the features' contributions over the temperatures range from 2 MK to 16 MK in 2 MK increments.

### A.1. Mg Dominated Features

There are four features in the DAXSS spectral range (0.5-10 keV) that have their emissions dominated by Mg lines: 1.34 keV, 1.47 keV, 1.57 keV, and 1.74 keV. None of these features, as listed in Table A.1, are dominated by Mg over the full 2 MK to 16 MK range, so at least two of these features need to be observed for modeling over the full temperature range.

The 1.34 keV feature is dominated by Mg XI lines for the 2 MK to 10 MK range and then becomes dominated by Fe XXI and Fe XXII lines for the 12 MK to 16 MK range.

The 1.47 keV feature is dominated by Mg XXII lines over the full temperature range, although this feature will be weak for the 2-4 MK cooler temperature range. The Fe XXIII lines start to affect the blend for a temperature of 12 MK and hotter.

The 1.57 keV feature is the other Mg feature sensitive to the lower temperature range from 2 MK to 10 MK. This feature has strong contributions from both Mg XI and Al XII emissions. This feature is not expected to be detected for temperatures in the 12 MK to 16 MK range.

The 1.74 keV feature is only detectable for the warmer temperature range of 8 MK to 16 MK and is mostly from Mg XII lines. This feature also has significant contributions from Al XIII over this same temperature range, and Fe XXIV lines begin to contribute more for 14 MK and hotter.

## A.2. Si Dominated Features

The three features between 1.85 keV and 2.17 keV are the most ideal set of features in the DAXSS spectra to study low-FIP abundance changes because those features are strongly dominated by Si lines over the full temperature range (2 MK to 16 MK) and with very minor blends from other elements. The 1.85 keV, 2.00 keV, and 2.17 keV features are listed in Table A.2. The 1.85 keV feature is the brightest (relative to the continuum) for the 2MK to 10MK temperature range, and the 1.85 keV and 2.00 keV features are about equally bright for 10 MK and hotter.

## A.3. Fe Dominated Features

There are four features between 0.71 keV and 1.34 keV that have strong Fe contributions, with three of those features listed in Table A.3 and one already listed in Table A.1 (Mg dominated features). There are also three more Fe-dominated features at higher energy between 6.57 keV and 8.19 keV as listed in Table A.4. These higher energy features are only detected in the DAXSS spectra during flare events. None of these Fe-dominated features have much emissions at 2 MK, so DAXSS SXR spectra are not appropriate for studying the Fe abundance for quiet-Sun conditions; however, combining DAXSS SXR and SDO/EVE EUV spectral observations can address the Fe abundance at the cooler coronal temperatures. We note that most of these Fe-dominated features have energy (wavelength) shifts with temperature changes due to either significant blends with other elements or the transition to higher ionized Fe at the hotter temperatures.

The 0.71 keV feature is dominated by O VIII at 2 MK and then again at 14 MK and 16 MK; these O emissions also peak near 0.65 keV. The Fe XVII and Fe XVIII emissions dominate between 4 MK and 8 MK near 0.73 keV. This feature and the 0.84 keV feature could be useful for studying Fe abundance for active region conditions.

The 0.84 keV feature is dominated by Ne IX at 2 MK and near 0.91 keV, and then Fe dominates for the 4 MK to 10 MK range. This feature is not expected to be detectable for temperatures above 10 MK. The primary Fe ions for this feature change with temperature, starting at Fe XVII for 4 MK, then Fe XVIII by 8 MK, and Fe XIX at 10 MK. These Fe emissions are all near 0.86 keV.

The 0.99 keV feature has a strong blend of Fe XVII and Ne X emissions for 4-6 MK and then Fe ions (Fe XX, Fe XXI, and Fe XXII) dominate for 12-16 MK range. The 0.84 keV and 0.99 keV features could be used together to cover the almost full range of temperature (4 MK to 16 MK).

The 1.34 keV feature with strong blends of Mg and Fe emissions was already discussed in Section A.1.

The higher energy Fe-dominated features usually only appear in DAXSS spectra during flares. That is, DAXSS spectra are typically limited to a high energy limit of 4-5 keV for non-flaring times. The 6.57 keV feature is the brighter feature and is detectable from 6 MK to 16 MK. There are many Fe emissions in this feature though, ranging from Fe XVIII to Fe XXV, so there are energy (wavelength) shifts with temperature changes for this feature. The 7.79 keV and 8.19 keV features are both from Fe XXV emission lines, and so their peak energy doesn't shift much, but they are weaker (hard-to-detect) features. Phillips (2004) discusses these higher-energy Fe features in the RHESSI spectra, but with a focus more on hotter temperatures of 20 MK and higher. Our results for the energy shift of the 6.57 keV with temperatures below 16 MK is not as large as what Phillips (2004) shows for an older version of CHIANTI modeling.

### A.4. S Dominated Features

The three features between 2.44 keV and 2.88 keV are dominated by S with just moderate blend of Si for two of them. S is a mid-FIP element, so its abundance changes are not expected to be as large as low-FIP elements for different solar conditions. The 2.44 keV, 2.61 keV, and 2.88 keV features are listed in Table A.5. The 2.44 keV feature is dominated by S XV lines over almost the full temperature range (6 MK to 16 MK). The 2.61 keV and 2.88 keV features are only detectable in the DAXSS spectra for the hotter temperatures above 12 MK. The 2.61 keV feature is dominated with S XVI lines and with moderate blend with Si XIV. The 2.88 keV feature is highly dominated by S XV lines.

### A.5. Ca Dominated Features

There are two Ca dominated features, one at 0.57 keV at the lower end of the DAXSS energy range and one at 3.88 keV. The 0.57 keV feature is only detected for cooler temperatures (2-6 MK) and has strong blend of O (O VII) and Ca (Ca XIII, Ca XIV, Ca XV, and Ca XVI). With larger uncertainty for the irradiance at the lowest energies in the DAXSS spectra, we usually use 0.7 keV as the lower limit for modeling the DAXSS spectra. The other Ca feature at 3.88 keV is excellent for modeling the Ca abundance but limited for warmer temperatures (6-16 MK) appropriate for active region and flare conditions. The 3.88 keV feature is dominated by Ca XIX lines as listed in Table A.6.

### A.6. Ar Dominated Feature

The 3.11 keV feature is the only Ar-dominated feature in the DAXSS spectra. Ar is a high-FIP element, so its abundance changes are expected to be small with different solar conditions. The 3.11 keV feature (listed in Table A.6) is from Ar XVII lines and with about 15% blend from S XV.

Table A.1. Mg Features in SXR Spectra: Brightest Three Elements and Ions

The features' energy peaks, ratios of features' intensity peak to continuum, and percent contribution by element/ion are listed for temperatures between 2 MK and 16 MK.

| Plasma Temp. | 1.34 keV Feature | | 1.47 keV Feature | | 1.57 keV Feature | | 1.74 keV Feature | |
|---|---|---|---|---|---|---|---|---|
| | Elements | Ions | Elements | Ions | Elements | Ions | Elements | Ions |
| 2MK | 1.341 keV, x4.0 | | ... | | 1.559 keV, x1.4 | | ... | |
| | Mg 99% | Mg XI 99% | | | Mg 75% | Mg XI 75% | | |
| | | | | | Al 24% | Al XII 24% | | |
| | | | | | Si 1% | Si XIII 1% | | |
| 4MK | 1.343 keV, x4.8 | | ... | | 1.582 keV, x2.0 | | ... | |
| | Mg 97% | Mg XI 97% | | | Mg 61% | Mg XI 60% | | |
| | Ne 1% | Ne X 1% | | | Al 38% | Al XII 38% | | |
| | Ni 1% | Ni XIX 1% | | | | | | |
| 6MK | 1.343 keV, x3.8 | | 1.467 keV, x1.8 | | 1.579 keV, x1.9 | | ... | |
| | Mg 94% | Mg XI 94% | Mg 94% | Mg XII 88% | Mg 59% | Mg XI 54% | | |
| | Ni 3% | Ni XIX 2% | Ni 4% | Mg XI 5% | Al 40% | Al XII 40% | | |
| | Fe 2% | Fe XIX 1% | Na 1% | Ni XIX 3% | Na 1% | Mg XII 5% | | |
| 8MK | 1.340 keV, x2.4 | | 1.469 keV, x2.7 | | 1.576 keV, x1.8 | | 1.739 keV, x1.1 | |
| | Mg 80% | Mg XI 79% | Mg 94% | Mg XII 91% | Mg 54% | Mg XI 45% | Mg 69% | Mg XII 51% |
| | Fe 14% | Fe XX 5% | Fe 3% | Ni XIX 2% | Al 44% | Al XII 44% | Al 21% | Al XIII 21% |
| | Ni 5% | Fe XXI 4% | Ni 2% | Mg XI 2% | Na 1% | Mg XII 9% | Si 10% | Mg XI 18% |
| 10MK | 1.329 keV, x1.8 | | 1.467 keV, x3.1 | | 1.567 keV, x1.6 | | 1.745 keV, x1.3 | |
| | Fe 48% | Mg XI 43% | Mg 83% | Mg XII 82% | Mg 46% | Al XII 40% | Mg 60% | Mg XII 50% |
| | Mg 43% | Fe XXI 23% | Fe 14% | Fe XXI 7% | Al 40% | Mg XI 32% | Al 29% | Al XIII 28% |
| | Ni 6% | Fe XX 10% | Ni 2% | Fe XXII 4% | Fe 11% | Mg XII 14% | Si 10% | Si XIII 10% |
| 12MK | 1.322 keV, x1.5 | | 1.464 keV, x2.7 | | ... | | 1.739 keV, x1.3 | |
| | Fe 71% | Fe XXI 40% | Mg 66% | Mg XII 66% | | | Mg 51% | Mg XII 46% |
| | Mg 21% | Mg XI 21% | Fe 32% | Fe XXI 10% | | | Al 33% | Al XIII 32% |
| | Ni 6% | Fe XXII 15% | Ni 1% | Fe XXIII 9% | | | Si 7% | Si XIII 7% |

*CONTINUED* Table A.1. Mg Features in SXR Spectra: Brightest Three Elements and Ions

The features' energy peaks, ratios of features' intensity peak to continuum, and percent contribution by element/ion are listed for temperatures between 2 MK and 16 MK.

| Plasma Temp. | 1.34 keV Feature Elements | Ions | 1.47 keV Feature Elements | Ions | 1.57 keV Feature Elements | Ions | 1.74 keV Feature Elements | Ions |
|---|---|---|---|---|---|---|---|---|
| 14MK | 1.344 keV, x1.2 | | 1.466 keV, x2.2 | | … | | 1.732 keV, x1.2 | |
| | Fe 74% | Fe XXII 32% | Mg 56% | Mg XII 55% | | | Mg 42% | Mg XII 38% |
| | Mg 19% | Fe XXI 28% | Fe 42% | Fe XXIII 17% | | | Al 31% | Al XIII 31% |
| | Ni 5% | Mg XI 19% | Ni 1% | Fe XXII 11% | | | Fe 18% | Fe XXIV 13% |
| 16MK | 1.347 keV, x1.2 | | 1.469 keV, x2.0 | | … | | 1.726 keV, x1.1 | |
| | Fe 71% | Fe XXII 35% | Mg 53% | Mg XII 52% | | | Mg 34% | Mg XII 32% |
| | Mg 17% | Fe XXI 22% | Fe 46% | Fe XXIII 20% | | | Al 28% | Al XIII 27% |
| | Ni 10% | Mg XI 17% | Ni 1% | Fe XXIV 12% | | | Fe 27% | Fe XXIV 21% |

Table A.2. Si Features in SXR Spectra: Brightest Three Elements and Ions

The features' energy peaks, ratios of features' intensity peak to continuum, and percent contribution by element/ion are listed for temperatures between 2 MK and 16 MK.

| Plasma Temp. | 1.85 keV Feature Elements | 1.85 keV Feature Ions | 2.00 keV Feature Elements | 2.00 keV Feature Ions | 2.17 keV Feature Elements | 2.17 keV Feature Ions |
|---|---|---|---|---|---|---|
| 2MK | 1.847 keV, x3.3 | | ... | | 2.136 keV, x1.6 | |
| | Si 99% | Si XIII 98% | | | Si 100% | Si XIII 100% |
| | Al 1% | Al XII 1% | | | | |
| 4MK | 1.853 keV, x4.1 | | ... | | 2.154 keV, x1.3 | |
| | Si 99% | Si XIII 99% | | | Si 99% | Si XIII 99% |
| | Al 1% | Al XII 1% | | | P 1% | P XIV 1% |
| 6MK | 1.854 keV, x4.4 | | 2.000 keV, x1.4 | | 2.178 keV, x1.6 | |
| | Si 97% | Si XIII 97% | Si 87% | Si XIV 86% | Si 98% | Si XIII 98% |
| | Mg 2% | Mg XII 2% | Al 12% | Al XII 10% | P 1% | P XIV 1% |
| | Al 1% | Al XII 1% | S 1% | Al XIII 1% | Al 1% | Al XIII 1% |
| 8MK | 1.855 keV, x4.4 | | 2.003 keV, x2.2 | | 2.181 keV, x1.6 | |
| | Si 96% | Si XIII 96% | Si 96% | Si XIV 95% | Si 97% | Si XIII 97% |
| | Mg 3% | Mg XII 3% | Al 4% | Al XII 2% | Al 1% | Al XIII 1% |
| | Al 1% | Al XII 1% | | Al XIII 1% | P 1% | P XIV 1% |
| 10MK | 1.856 keV, x3.7 | | 2.004 keV, x2.8 | | 2.181 keV, x1.6 | |
| | Si 96% | Si XIII 96% | Si 98% | Si XIV 98% | Si 96% | Si XIII 96% |
| | Mg 3% | Mg XII 3% | Al 2% | Al XII 1% | Al 2% | Al XIII 2% |
| | Al 1% | Al XII 1% | | | P 1% | P XIV 1% |
| 12MK | 1.856 keV, x3.0 | | 2.005 keV, x3.2 | | 2.182 keV, x1.5 | |
| | Si 96% | Si XIII 95% | Si 99% | Si XIV 98% | Si 96% | Si XIII 95% |
| | Mg 4% | Mg XII 4% | Al 1% | Al XIII 1% | Al 3% | Al XIII 3% |
| | Al 1% | Al XII 1% | | | P 1% | P XIV 1% |

*CONTINUED* Table A.2. Si Features in SXR Spectra: Brightest Three Elements and Ions

The features' energy peaks, ratios of features' intensity peak to continuum, and percent contribution by element/ion are listed for temperatures between 2 MK and 16 MK.

| Plasma Temp. | 1.85 keV Feature | | 2.00 keV Feature | | 2.17 keV Feature | |
|---|---|---|---|---|---|---|
| | Elements | Ions | Elements | Ions | Elements | Ions |
| 14MK | 1.856 keV, x2.4 | | 2.005 keV, x3.3 | | 2.182 keV, x1.4 | |
| | Si 95% | Si XIII 94% | Si 99% | Si XIV 99% | Si 94% | Si XIII 94% |
| | Mg 4% | Mg XII 4% | Al 1% | Al XIII 1% | Al 4% | Al XIII 4% |
| | Al 1% | Al XII 1% | | | P 2% | P XIV 2% |
| 16MK | 1.856 keV, x2.1 | | 2.005 keV, x3.2 | | 2.182 keV, x1.3 | |
| | Si 93% | Si XIII 92% | Si 99% | Si XIV 99% | Si 93% | Si XIII 93% |
| | Mg 5% | Mg XII 5% | Al 1% | Al XIII 1% | Al 5% | Al XIII 4% |
| | Fe 1% | Fe XXIV 1% | | | P 2% | P XIV 2% |

Table A.3. Low-Energy Fe Features in SXR Spectra: Brightest Three Elements and Ions

The features' energy peaks, ratios of features' intensity peak to continuum, and percent contribution by element/ion are listed for temperatures between 2 MK and 16 MK.

| Plasma Temp. | 0.71 keV Feature | | 0.84 keV Feature | | 0.99 keV Feature | |
|---|---|---|---|---|---|---|
| | Elements | Ions | Elements | Ions | Elements | Ions |
| 2MK | 0.647 keV, x3.7 | | 0.909 keV, x4.6 | | … | |
| | O 95% | O VIII 65% | Ne 91% | Ne IX 91% | | |
| | Fe 2% | O VII 30% | Fe 7% | Fe XVII 6% | | |
| | Cr 1% | Fe XVII 1% | O 2% | O VIII 1% | | |
| 4MK | 0.730 keV, x3.0 | | 0.815 keV, x3.4 | | 1.014 keV, x2.2 | |
| | Fe 94% | Fe XVII 91% | Fe 95% | Fe XVII 85% | Fe 62% | Fe XVII 59% |
| | O 4% | O VIII 4% | O 4% | Fe XVIII 9% | Ne 25% | Ne X 21% |
| | Ca 1% | Fe XVIII 3% | Ni 1% | O VIII 4% | Ni 12% | Ni XIX 12% |
| 6MK | 0.735 keV, x5.7 | | 0.820 keV, x2.2 | | 1.013 keV, x1.8 | |
| | Fe 98% | Fe XVII 83% | Fe 97% | Fe XVII 63% | Fe 65% | Fe XVII 50% |
| | O 1% | Fe XVIII 13% | O 1% | Fe XVIII 29% | Ne 18% | Ne X 17% |
| | | Fe XIX 1% | Ni 1% | Fe XIX 5% | Ni 16% | Ni XIX 13% |
| 8MK | 0.739 keV, x5.3 | | 0.859 keV, x1.9 | | … | |
| | Fe 98% | Fe XVII 67% | Fe 95% | Fe XVIII 45% | | |
| | O 1% | Fe XVIII 23% | Ni 3% | Fe XIX 23% | | |
| | | Fe XIX 7% | | Fe XVII 21% | | |
| 10MK | … | | 0.893 keV, x1.5 | | … | |
| | | | Fe 95% | Fe XIX 41% | | |
| | | | Ni 4% | Fe XVIII 25% | | |
| | | | Cr 1% | Fe XX 17% | | |
| 12MK | … | | … | | 0.943 keV, x1.5 | |
| | | | | | Fe 95% | Fe XX 36% |
| | | | | | Ni 4% | Fe XIX 24% |
| | | | | | | Fe XXI 24% |

*CONTINUED* Table A.3. Low-Energy Fe Features in SXR Spectra: Brightest Three Elements and Ions
The features' energy peaks, ratios of features' intensity peak to continuum, and percent contribution by element/ion are listed for temperatures between 2 MK and 16 MK.

| Plasma Temp. | 0.71 keV Feature Elements | 0.71 keV Feature Ions | 0.84 keV Feature Elements | 0.84 keV Feature Ions | 0.99 keV Feature Elements | 0.99 keV Feature Ions |
|---|---|---|---|---|---|---|
| 14MK | 0.653 keV, x1.5 | | … | | 1.010 keV, x1.6 | |
| | O 74% | O VIII 74% | | | Fe 94% | Fe XXI 37% |
| | Ca 17% | Ca XVIII 16% | | | Ni 4% | Fe XXII 27% |
| | Fe 6% | Fe XVIII 3% | | | Ne 2% | Fe XXIII 14% |
| 16MK | 0.651 keV, x1.8 | | … | | 1.030 keV, x1.6 | |
| | O 76% | O VIII 76% | | | Fe 94% | Fe XXII 34% |
| | Ca 19% | Ca XVIII 17% | | | Ni 3% | Fe XXIII 28% |
| | Fe 2% | Ca XVII 1% | | | Ne 3% | Fe XXI 24% |

Table A.4. High-Energy Fe Features in SXR Spectra: Brightest Three Elements and Ions

The features' energy peaks, ratios of features' intensity peak to continuum, and percent contribution by element/ion are listed for temperatures between 2 MK and 16 MK.

| Plasma Temp. | 6.57 keV Feature | | 7.79 keV Feature | | 8.19 keV Feature | |
|---|---|---|---|---|---|---|
| | Elements | Ions | Elements | Ions | Elements | Ions |
| 2MK | … | | … | | … | |
| 4MK | … | | … | | … | |
| 6MK | 6.430 keV, x1.4 | | … | | … | |
| | Fe 100% | Fe XVIII 88% | | | | |
| | | Fe XIX 10% | | | | |
| | | Fe XX 1% | | | | |
| 8MK | 6.446 keV, x1.3 | | … | | … | |
| | Fe 100% | Fe XVIII 62% | | | | |
| | | Fe XXI 13% | | | | |
| | | Fe XX 12% | | | | |
| 10MK | 6.554 keV, x1.2 | | … | | … | |
| | Fe 100% | Fe XXIII 35% | | | | |
| | | Fe XXI 21% | | | | |
| | | Fe XXII 20% | | | | |
| 12MK | 6.607 keV, x1.4 | | 7.789 keV, x1.1 | | … | |
| | Fe 100% | Fe XXIII 33% | Fe 99% | Fe XXV 99% | | |
| | | Fe XXIV 29% | Ni 1% | Ni XXVII 1% | | |
| | | Fe XXV 27% | | | | |
| 14MK | 6.633 keV, x1.6 | | 7.792 keV, x1.2 | | 8.185 keV, x1.1 | |
| | Fe 100% | Fe XXV 54% | Fe 95% | Fe XXV 86% | Fe 100% | Fe XXV 100% |
| | | Fe XXIV 27% | Ni 5% | Ni XXVII 5% | | |
| | | Fe XXIII 16% | | | | |

*CONTINUED* Table A.4. High-Energy Fe Features in SXR Spectra: Brightest Three Elements and Ions
The features' energy peaks, ratios of features' intensity peak to continuum, and percent contribution by element/ion are listed for temperatures between 2 MK and 16 MK.

| Plasma Temp. | 6.57 keV Feature Elements | 6.57 keV Feature Ions | 7.79 keV Feature Elements | 7.79 keV Feature Ions | 8.19 keV Feature Elements | 8.19 keV Feature Ions |
|---|---|---|---|---|---|---|
| 16MK | 6.646 keV, x1.6 | | 7.794 keV, x1.4 | | 8.188 keV, x1.2 | |
| | Fe 100% | Fe XXV 70% | Fe 89% | Fe XXV 89% | Fe 100% | Fe XXV 100% |
| | | Fe XXIV 21% | Ni 11% | Ni XXVII 11% | | |
| | | Fe XXIII 8% | | | | |

Table A.5. S Features in SXR Spectra: Brightest Three Elements and Ions

The features' energy peaks, ratios of features' intensity peak to continuum, and percent contribution by element/ion are listed for temperatures between 2 MK and 16 MK.

| Plasma Temp. | 2.44 keV Feature | | 2.61 keV Feature | | 2.88 keV Feature | |
|---|---|---|---|---|---|---|
| | Elements | Ions | Elements | Ions | Elements | Ions |
| 2MK | … | | … | | … | |
| 4MK | 2.442 keV, x1.9 | | … | | … | |
| | S 98% | S XIV 98% | | | | |
| | Si 2% | Si XIII 2% | | | | |
| 6MK | 2.445 keV, x2.0 | | … | | … | |
| | S 97% | S XV 96% | | | | |
| | Si 3% | Si XIV 2% | | | | |
| | | Si XIII 1% | | | | |
| 8MK | 2.444 keV, x2.0 | | … | | … | |
| | S 92% | S XV 92% | | | | |
| | Si 8% | Si XIV 6% | | | | |
| | | Si XIII 2% | | | | |
| 10MK | 2.443 keV, x1.9 | | … | | … | |
| | S 87% | S XV 86% | | | | |
| | Si 13% | Si XIV 11% | | | | |
| | | Si XIII 2% | | | | |
| 12MK | 2.442 keV, x1.8 | | … | | 2.876 keV, x1.2 | |
| | S 82% | S XV 82% | | | S 97% | S XV 97% |
| | Si 18% | Si XIV 16% | | | Cl 3% | Cl XVI 2% |
| | | Si XIII 2% | | | | |

*CONTINUED* Table A.5. S Features in SXR Spectra: Brightest Three Elements and Ions

The features' energy peaks, ratios of features' intensity peak to continuum, and percent contribution by element/ion are listed for temperatures between 2 MK and 16 MK.

| Plasma Temp. | 2.44 keV Feature | | 2.61 keV Feature | | 2.88 keV Feature | |
|---|---|---|---|---|---|---|
| | Elements | Ions | Elements | Ions | Elements | Ions |
| 14MK | 2.440 keV, x1.6 | | 2.608 keV, x1.4 | | 2.878 keV, x1.2 | |
| | S 79% | S XV 79% | S 82% | S XVI 82% | S 97% | S XV 97% |
| | Si 21% | Si XIV 19% | Si 17% | Si XIV 17% | Cl 3% | Cl XVI 2% |
| | | Si XIII 1% | | | | |
| 16MK | 2.440 keV, x1.6 | | 2.612 keV, x1.5 | | 2.879 keV, x1.2 | |
| | S 77% | S XV 77% | S 87% | S XVI 87% | S 96% | S XV 96% |
| | Si 23% | Si XIV 22% | Si 12% | Si XIV 12% | Cl 3% | Cl XVI 2% |
| | | Si XIII 1% | | | | |

Table A.6. Ca and Ar Features in SXR Spectra: Brightest Three Elements and Ions

The features' energy peaks, ratios of features' intensity peak to continuum, and percent contribution by element/ion are listed for temperatures between 2 MK and 16 MK.

| Plasma Temp. | Ca 0.56 keV Feature | | Ca 3.88 keV Feature | | Ar 3.11 keV Feature | |
|---|---|---|---|---|---|---|
| | Elements | Ions | Elements | Ions | Elements | Ions |
| 2MK | 0.568 keV, x3.8 | | ... | | ... | |
| | O 95% | O VII 94% | | | | |
| | N 2% | N VII 2% | | | | |
| | Ca 2% | Ca XI 1% | | | | |
| 4MK | 0.462 keV, x1.2 | | ... | | ... | |
| | Ca 63% | Ca XIII 25% | | | | |
| | Ar 17% | Ca XIV 21% | | | | |
| | N 5% | Ca XV 7% | | | | |
| 6MK | 0.555 keV, x1.1 | | 3.874 keV, x1.4 | | 3.108 keV, x1.3 | |
| | Ca 65% | Ca XVI 24% | Ca 95% | Ca XIX 95% | Ar 81% | Ar XVII 81% |
| | O 16% | Ca XV 21% | Ar 5% | Ar XVII 5% | S 18% | S XV 18% |
| | Cr 5% | O VII 16% | | | | |
| 8MK | ... | | 3.878 keV, x1.5 | | 3.112 keV, x1.3 | |
| | | | Ca 98% | Ca XIX 98% | Ar 83% | Ar XVII 83% |
| | | | Ar 2% | Ar XVII 2% | S 17% | S XV 16% |
| | | | | | | S XVI 1% |
| 10MK | ... | | 3.882 keV, x1.5 | | 3.115 keV, x1.3 | |
| | | | Ca 99% | Ca XIX 98% | Ar 81% | Ar XVII 81% |
| | | | Ar 1% | Ar XVII 1% | S 19% | S XV 15% |
| | | | | | | S XVI 4% |
| 12MK | ... | | 3.884 keV, x1.6 | | 3.115 keV, x1.3 | |
| | | | Ca 99% | Ca XIX 99% | Ar 79% | Ar XVII 78% |
| | | | Ar 1% | Ar XVII 1% | S 21% | S XV 14% |
| | | | | | | S XVI 7% |

*CONTINUED* Table A.6. Ca and Ar Features in SXR Spectra: Brightest Three Elements and Ions

The features' energy peaks, ratios of features' intensity peak to continuum, and percent contribution by element/ion are listed for temperatures between 2 MK and 16 MK.

| Plasma Temp. | Ca 0.56 keV Feature | | Ca 3.88 keV Feature | | Ar 3.11 keV Feature | |
|---|---|---|---|---|---|---|
| | Elements | Ions | Elements | Ions | Elements | Ions |
| 14MK | … | | 3.885 keV, x1.6 | | 3.117 keV, x1.3 | |
| | | | Ca 99% | Ca XIX 99% | Ar 77% | Ar XVII 76% |
| | | | Ar 1% | Ar XVII 1% | S 23% | S XV 12% |
| | | | | | | S XVI 11% |
| 16MK | … | | 3.887 keV, x1.6 | | 3.117 keV, x1.3 | |
| | | | Ca 99% | Ca XIX 99% | Ar 74% | Ar XVII 74% |
| | | | Ar 1% | Ar XVII 1% | S 26% | S XVI 15% |
| | | | | | | S XV 10% |